\documentstyle[aps,prb,twocolumn,psfig]{revtex}

\newcommand{\U}{UPt$_3$~}
\newcommand{\vq}{\bf q\rm~}
\newcommand{\vk}{\bf k\rm~}

\newcommand{\vH}{\bf H\rm~}
\newcommand{\vv}{\bf v\rm~}
\newcommand{\vd}{\bf d\rm~}

\newcommand{\la}{\langle}
\newcommand{\ra}{\rangle}
\newcommand{\De}{$\Delta(\bf k\rm)~$}

\begin{document}

\draft

\title{Thermal conductivity in B- and C- phase of \U}

\author{ P. Thalmeier$^1$ and K. Maki$^{2,3}$}

\address{$^1$Max-Planck-Institute for the Chemical Physics of Solids,
N\"othnitzer Str.40, 01187 Dresden, Germany}
\address{$^2$Max-Planck-Institute for the Physics of Complex Systems, 
N\"othnitzer Str.38, 01187 Dresden, Germany}
\address{$^3$Department of Physics and Astronomy,
University of Southern California, Los Angeles,}
\address{CA 90089-0484, USA}

\maketitle

\begin{abstract}
Although the superconductivity in \U is one of the most well studied,
there are still lingering questions about the nodal directions in the B
and C phase in the presence of a magnetic field. Limiting ourselves to
the low temperature regime (T$\ll\Delta(0)$), we study the
magnetothermal conductivity with in semiclassical approximation using
Volovik's approach. The angular dependence of the magnetothermal
conductivity for an arbitrary field direction is calculated for both
phases. We show that cusps in the polar angle dependence appear in B and
C phases which are due to the polar point nodes.
\end{abstract}

\pacs{PACS numbers: 74.20.Rp, 74.25.Fy, 74.70.Tx} 

\section{Introduction}
Perhaps due to the exceptional appearance of two superconducting(sc)
transitions with two associated critical field curves, \U is one of
the most well studied Heavy Fermion superconductors. The history of
this subject was described in Ref.\cite{Joynt02}. The
T linear dependence of the low temperature thermal conductivity  both
parallel to the c axis and the a axis was in favor of the E$_{2u}$
rather than E$_{1g}$- type sc order
parameter\cite{Lussier96,Norman96,Graf96}. Further Pt- NMR experiments
confirmed the triplet
nature of the superconductivity and the details of the triplet gap
function \vd(\bf k\rm) n A, B, and C- phases were
determined\cite{Tou96,Tou98}. This assignment of the spin
configuration is consistent with the weak spin-orbit coupling
limit\cite{Tou98,Kitaoka00,Machida99} also assumed here. H$_{c2}$
along the c axis requires a strong Pauli limiting
effect\cite{Yang99}. This favors strong spin orbit
coupling but then H$_{c2}$ results are inconsistent with the NMR
data. We assume that \vd(\bf k\rm)=$\Delta(\bf k\rm)\hat{z}$ to be
weakly pinned along c. Then the E$_{2u}$ gap functions of 
for the low temperature B and C phases of \U are given by

\begin{eqnarray}
\label{OP}
B: \Delta(\vartheta,\varphi)&=&
\frac{3}{2}\sqrt{3}\Delta\cos\vartheta\sin^2\vartheta\exp(\pm 2i\varphi)
\nonumber\\
C: \Delta(\vartheta,\varphi)&=&
\frac{3}{2}\sqrt{3}\Delta\cos\vartheta\sin^2\vartheta\cos(2\varphi)
\end{eqnarray} 

\begin{figure}
\centerline{\psfig{figure=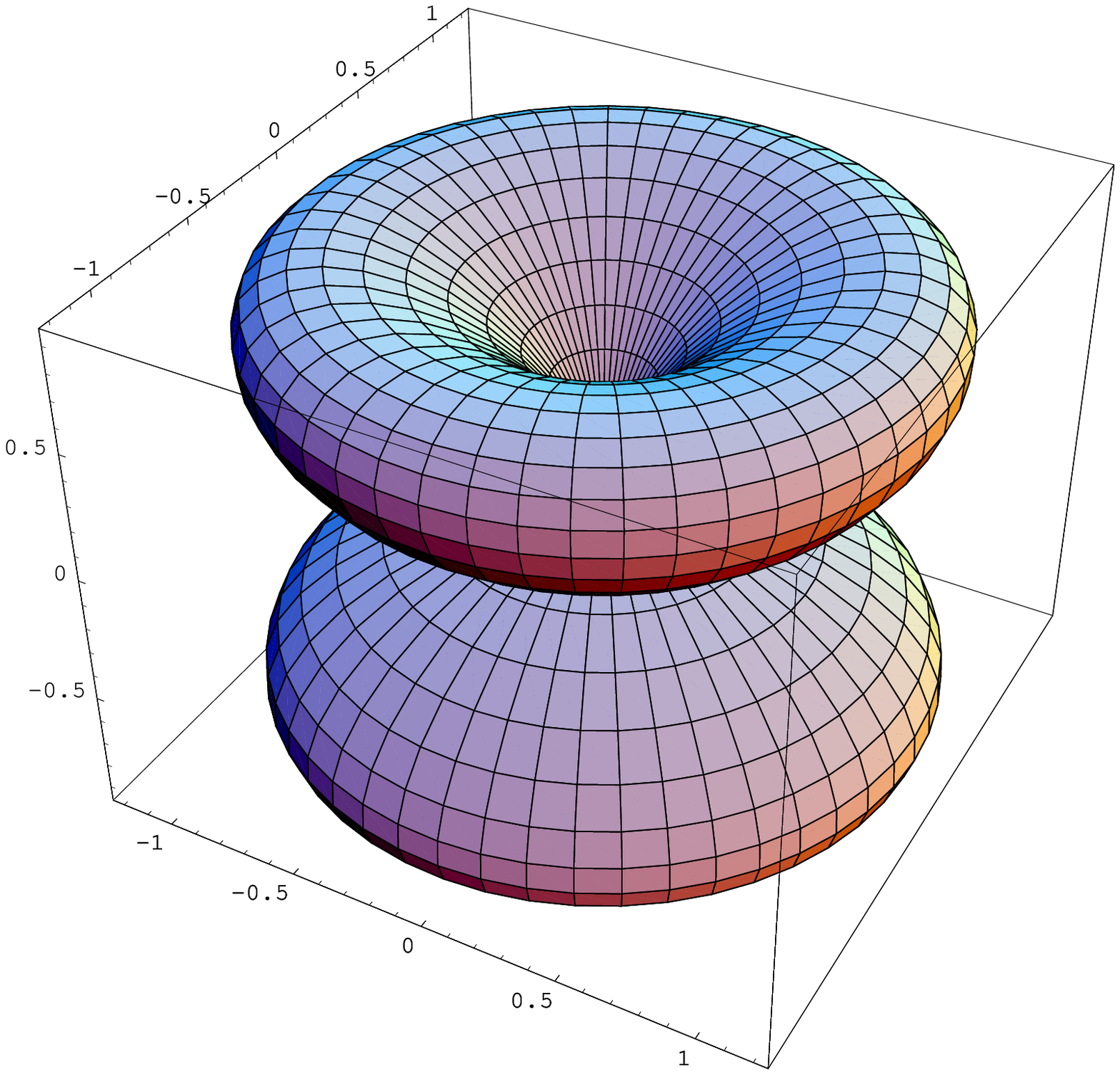,width=.48\columnwidth}\hfill
\psfig{figure=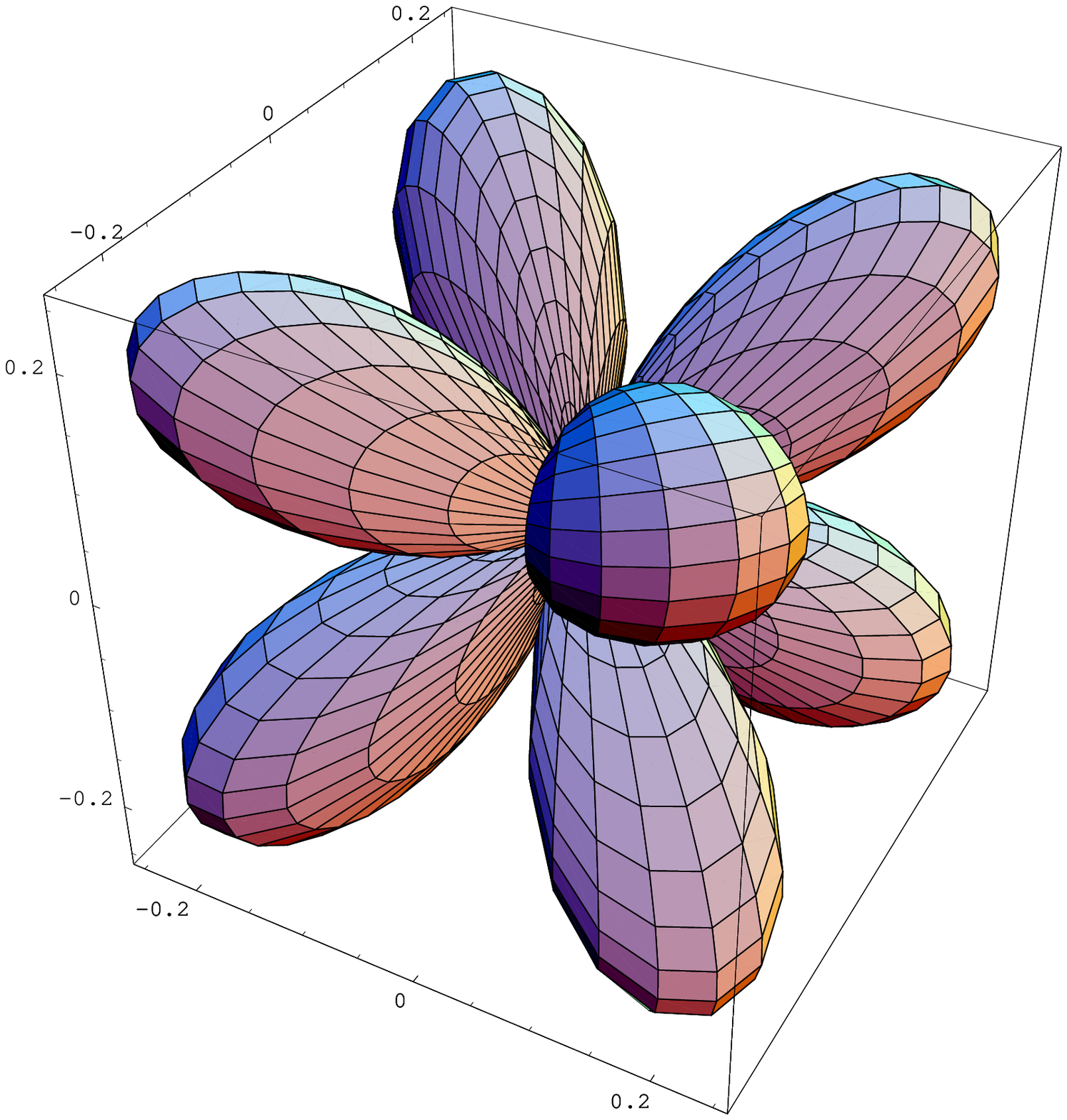,width=.48\columnwidth}}
\vspace{1cm}
\caption{Spherical plots of $|\Delta(\vartheta,\varphi)|$ for B phase (left)
and C phase (right). For the C- phase two additional vertical nodal
planes appear
an angle $\frac{\pi}{4}$ away from the vertical plane which contains
\vH assumed at $\phi=\frac{\pi}{4}$.}
\label{GAPFUN}
\end{figure}

Here $\vartheta, \varphi$ are the polar and azimuthal angles of
\vk. The angular
dependence of $\Delta(\vartheta,\varphi)$ is shown in
Fig.~\ref{GAPFUN}. In the B- phase the poles are second order node
points and the equator is a node line. In the C- phase two additional
vertical nodal lines appear at 45 degrees away from the vertical plane
containing \vH. The thermal conductivity in the vortex phase for field
along the symmetry axis also decided in favor of the
E$_{2u}$ state and against the E$_{1g}$- state\cite{Houssa97}, this is
reviewed in \cite{Brison00}. The angular dependent thermal
conductivity has been measured in \cite{Suderow97} and analyzed in
\cite{Maki00}. Unfortunately the experiment was done for T$>$0.3 K
which is not sufficiently low to determine node structures. Recently
we have been studying the thermal conductivity in nodal
superconductors\cite{Won00,Dahm00,Won01b,Won01c,Thalmeier02,Maki02}. In
particular we shall restrict to the low temperature
limit T$\ll\tilde{v}\sqrt{eH}\ll\Delta(0)$ and the superclean limit
$(\Gamma\Delta)^\frac{1}{2}\ll\tilde{v}\sqrt{eH}$, where
$\tilde{v}=(v_av_c)^\frac{1}{2}$, $\Gamma$ is the scattering rate and
$v_{a,c}$ are the anisotropic Fermi velocities. The condition 
$\tilde{v}\sqrt{eH}\ll\Delta(0)$ can be satisfied in the B- phase while
$\tilde{v}\sqrt{eH}\leq\Delta(0)$ in the C- phase since the latter
appears only for H$\geq$ 0.6 T and  H$\geq$ 1.2 T for field
along a and c respectively. Our results in the C-
phase may therefore only have qualitative significance. Restriction
to the superclean limit does not influence the main conclusions on the
connection between node topology and magnetothermal conductivity.
We first examine the quasiparticle DOS and the thermodynamic properties
of the B and C phase of \U in the presence of a
magnetic field with arbitrary orientation at low temperatures. 
Then we study the thermal conductivity in the B
and C phase in the low temperature regime which provides clear
evidence for the nodal positions in \De. In this regime the influence
of the AF order in \U may be neglected contrary to the A- phase regime. 

\section{Density of states and thermodynamics}

The quasiparticle DOS in the vortex state of \U is given by

\begin{equation}
\label{DENS}
g(0)=\it Re\rm\Biggl\la\frac{C_0-ix}
{\sqrt{(C_0-ix)^2+f^2}}\Biggr\ra
\end{equation} 

Where the brackets denote both Fermi surface and vortex lattice
average. With the form factor f$_B$(z)=$\frac{3\sqrt{3}}{2}(1-z^2)z$
and  f$_C$(z,z')=(2z'$^{2}$-1)f$_B$(z) (z=$\cos\vartheta$,
z'=$\cos\varphi$) for B and C- phase respectively the average may be
computed and one obtains

\begin{eqnarray}
\label{DENSEX}
g_B(0)&=&\frac{1}{\sqrt{3}}
\Bigl(\frac{\pi}{2}\la x\ra+
C_0\la\bigl[\ln(\frac{C_0}{x})-1\bigr ]\ra\Bigr)\nonumber\\
g_C(0)&=&\frac{1}{\sqrt{3}}
\Bigl(\frac{\pi}{2}\la x\ln\bigl(\frac{2}{x}\bigr)\ra +
\frac{2}{\pi}C_0\la\ln^2(\frac{2}{x})\ra\Bigr)
\end{eqnarray}

where x=$\frac{|\vv\cdot\vq|}{\Delta}$ is the normalized Doppler shift of
quasiparticle energies, C$_0= \lim_{\omega\rightarrow
0}Im(\tilde{\omega}/\Delta)$ and $\tilde{\omega}$ is the renormalized
quasiparticle energy in the presence of impurity
scattering\cite{Won00,Yang01}. Furthermore \vv is the quasiparticle
velocity and 2\vq is the pair momentum around
the vortex. Following Volovik\cite{Volovik93} we obtain

\begin{eqnarray}
\label{XAV}
\la x\ra_B&=&\frac{2}{\pi}\tilde{v}\frac{\sqrt{eH}}
{\Delta}I_B(\theta)\nonumber\\
\la x\ln(\frac{2}{x})\ra_C&=&
\frac{2}{\pi}\tilde{v}\frac{\sqrt{eH}}{\Delta}I_C(\theta)
\ln\bigl(\frac{\Delta}{\tilde{v}\sqrt{eH}}\bigr)\nonumber\\
I_B(\theta)&=&\alpha\sin\theta +\frac{2}{\pi}E(\sin\theta)\\
I_C(\theta)&=&I_B(\theta)+I(\theta)\nonumber
\end{eqnarray}

Here E($\sin\theta$) is the complete elliptic integral of the second
kind. Furthermore we used the definition

\begin{eqnarray}
I(\theta)&=&\frac{1}{2}\{\frac{f(\theta)}{\sqrt{|1-\alpha^2\sin^2\theta|}}
+2\alpha\sin\theta\nonumber\\
&&+\cos\theta(\frac{\pi}{2}-\tan^{-1}(\alpha\tan\theta))\}\\
f(\theta)&=&\left\{
\begin{array}{rl}
\cos^{-1}(\alpha\sin\theta) & \mbox{for $\sqrt{\alpha}\sin(\theta)<1$}\\
\cosh^{-1}(\alpha\sin\theta) & \mbox{for $\sqrt{\alpha}\sin(\theta)>1$}
\end{array}\right.\nonumber
\end{eqnarray}

where $\alpha=v_c/v_a$ is the anisotropy of Fermi velocities and
$\theta$ is the polar angle of \vH with respect to c
axis. The averages in Eq.~\ref{XAV} are evaluated by replacing the
\vk space integration by a summation over nodal positions and then
integrating out the superfluid velocity
field\cite{Maki00,Won00}. We assume a square vortex lattice for
simplicity, a hexagonal lattice would lead to an additional numerical
factor 0.93.

Substituting Eq.~\ref{XAV} into Eq.~\ref{DENSEX} we obtain

\begin{eqnarray}
g_{B}&=&\frac{1}{\sqrt{3}}\tilde{v}\frac{\sqrt{eH}}
{\Delta}I_{B}(\theta)\nonumber\\
g_{C}&=&\frac{2}{\pi\sqrt{3}}\tilde{v}\frac{\sqrt{eH}}
{\Delta}I_{C}(\theta)
\ln\bigl(\frac{\Delta}{\tilde{v}\sqrt{eH}}\bigr)
\end{eqnarray}

In addition we have C$_0$=$\frac{\Gamma}{\Delta}[g(0)]^{-1}$. The low
temperature specific heat, the spin susceptibility etc. are given by 

\begin{eqnarray}
\frac{C_s}{\gamma_NT}&=&
\frac{\chi_s}{\chi_N} = 1-\frac{\rho_s(H)}{\rho_s(0)}= g(0) 
\end{eqnarray}

where $\rho_s$ is the superfluid density. The $\theta$- dependence of
g(0) for B- and C- phases is shown in Figs.~\ref{BTHETA} and
\ref{CTHETA} respectively. 

\section{Thermal conductivity in the vortex phase}

The thermal conductivity tensor in the vortex phase depends on the
angles ($\theta,\phi$) of \vH  due to the angle dependence of
the Doppler shift energy $\Delta$x. Here $\phi$ is the azimuthal angle between
\vH and the direction of the heat current \bf j$_Q$ \rm in the ab- plane.
Following\cite{Maki02} we obtain for the B- phase

\begin{eqnarray} 
\frac{\kappa_{zz}}{\kappa_n}&=&\frac{2}{3}
\frac{v_av_c}{\Delta^2}(eH)I_B(\theta)F_B^{zz}(\theta) \nonumber\\
\frac{\kappa_{xx}}{\kappa_n}&=&\frac{1}{3}
\frac{v_a^2}{\Delta^2}(eH)I_B(\theta)F_B^{xx}(\theta)\nonumber\\
F_B^{zz}(\theta)&=&\sin\theta\\
F_B^{xx}(\theta,\phi)&=&\frac{2}{\pi}
\Bigl[\sin^2\phi E(\sin\theta)+\cos(2\phi)\frac{1}{3\sin^2\theta}\nonumber\\&&
\bigl(\cos^2\theta K(\sin\theta)-cos(2\theta)E(\sin\theta)\bigr)\Bigr]
\nonumber
\end{eqnarray}

\begin{figure}
\centerline{\psfig{figure=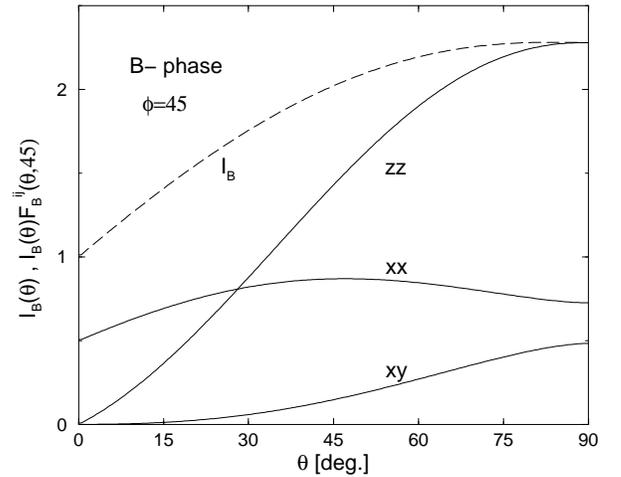,width=0.9\columnwidth}}
\vspace{0.5cm}
\caption{Polar field angle dependence of I$_B(\theta)$ 
and I$_B(\theta)$F$_B^{ij}(\theta,45)$ (ij= xx, zz, xy)
which determine the $\theta$- dependence of DOS g(0), thermal conductivities
($\kappa_{xx},\kappa_{zz}$)) and thermal Hall coefficient ($\kappa_{xy}$)
respectively.}
\label{BTHETA}
\end{figure}

The thermal Hall coefficient in the B- phase is obtained as

\begin{eqnarray}
\frac{\kappa_{xy}}{\kappa_n}(\theta)
&=&-\frac{v_a^2(eH)}{3\Delta^2}I_B(\theta)F_B^{xy}(\theta,\phi)
\nonumber\\
F_B^{xy}(\theta,\phi)&=&\frac{2}{\pi}\frac{\sin(2\phi)}{3\sin^2\theta}\\
&&\bigl[(2-\sin^2\theta)E(\sin\theta)-2\cos^2\theta
K(\sin\theta)\bigr]\nonumber
\end{eqnarray} 

\begin{figure}
\centerline{\psfig{figure=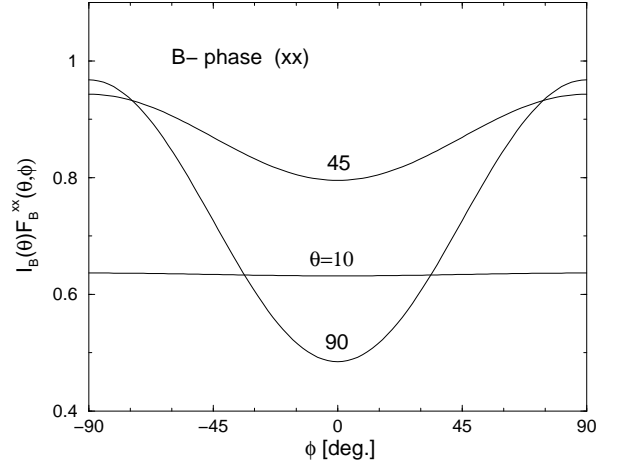,width=0.9\columnwidth}}
\vspace{0.5cm}
\caption{Azimuthal field angle dependence of
I$_B(\theta)$F$_B^{xx}(\theta,\phi)$ for various $\theta$. For
$\theta\rightarrow$0 the $\phi$ dependence is suppressed completely. The zz-
component is always $\phi$- independent.}
\label{BPHI}
\end{figure}

The $\theta$- and $\phi$- angle dependences of $\kappa_{ij}$ (ij=xx,zz,xy)
in the B- phase are shown in Fig.~\ref{BTHETA} and  Fig.~\ref{BPHI}. For heat
current along c ($\kappa_{zz}$) no $\phi$- dependence appears.

In the limit
$\theta=\frac{\pi}{2}$, I$_B(\frac{\pi}{2})$=$\alpha+\frac{2}{\pi}$ and then

\begin{eqnarray}
\kappa_{xx}\sim
\frac{1}{\pi}\bigl(1-\frac{1}{3}\cos(2\phi)\bigr)
;\;\;\;\;\kappa_{xy}&\sim& \frac{2}{3\pi}\sin(2\phi)
\end{eqnarray} 

The maximum in $\kappa_{xx}(90,\phi)$ at $\phi$= $\pm$90 occurs
for heat current $\perp$ \bf H \rm when the Doppler shift gives rise
to the largest quasiparticle DOS parallel to the heat current and we have
$\kappa_{xx}(\phi=90)/ \kappa_{xx}(\phi=0)$=2.

\begin{figure}
\centerline{\psfig{figure=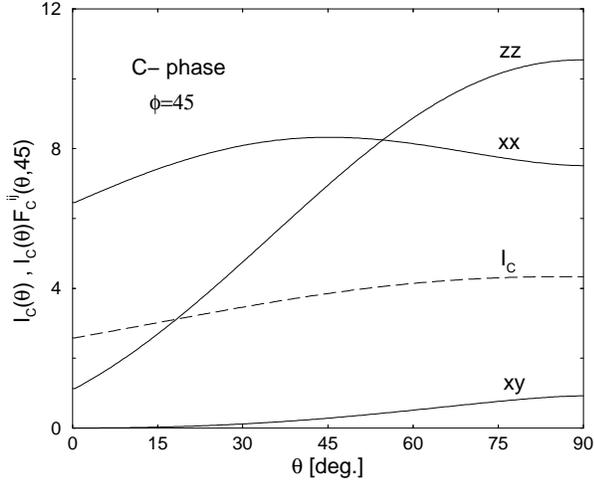,width=0.9\columnwidth}}
\vspace{0.5cm}
\caption{Polar field angle dependence of I$_C(\theta)$ 
and I$_C(\theta)$F$_C^{ij}(\theta,45)$ (ij= xx, zz, xy)}
\label{CTHETA}
\end{figure}

\begin{figure}
\centerline{\psfig{figure=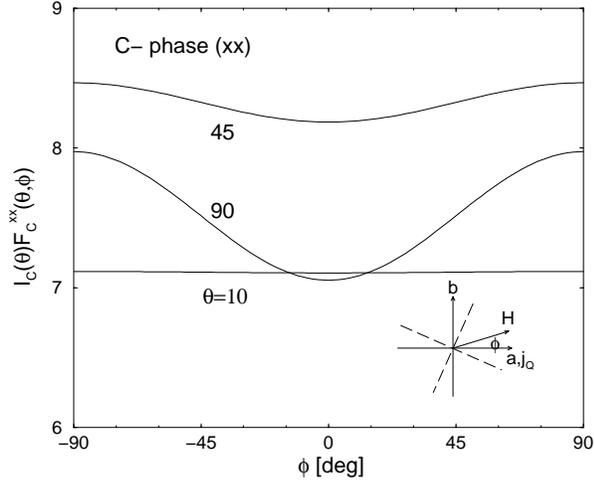,width=0.9\columnwidth}}
\vspace{0.5cm}
\caption{Azimuthal field angle dependence of
I$_C(\theta)$F$_C^{xx}(\theta,\phi)$ for various $\theta$. The zz-
component is again $\phi$-
independent. The inset shows the field and node geometry in the C-
phase. The node lines along c are lying in the planes perpendicular to
ab (dashed lines) which are mutually orthogonal. The field \vH lies at
half angle ($\frac{\pi}{4}$) in between forming an angle $\phi$ with the
heat current \bf j$_Q$ \rm along the a- axis. This geometry is
preserved for any $\phi$ because the nodal planes co- rotate with the field.}
\label{CPHI}
\end{figure}

Now we consider the C- phase. Again
$\kappa_{zz}$ does not exhibit a $\phi$- dependence. The C-phase
according to Eq.~\ref{GAPFUN} has two additional perpendicular node
lines. Rotating the field at a given
$\theta$ around c (i.e. changing $\phi$) will lead to a co- rotation
of these node lines such that the vertical plane containing \vH always
stays at half angle between the two perpendicular planes of the node
lines parallel to c\cite{Yang99}(see inset of Fig.~\ref{CPHI}). Consequently
$\kappa_{zz}(\theta)$ will again be independent of $\phi$ while
$\kappa_{xx}(\theta,\phi)$ depends on both field angles. We find for
heat current along c:

\begin{eqnarray} 
\frac{\kappa_{zz}}{\kappa_n}&=&
\frac{1}{6}\frac{v_a^2}{\Delta^2}(eH)I_C(\theta)F_C^{zz}(\theta)
\ln^2\bigl(\frac{\Delta}{\tilde{v}\sqrt{eH}}\bigr)\nonumber\\
F_C^{zz}(\theta)&=&\alpha\sin\theta
+\frac{2}{\pi}\int_{-1}^{1}dz|z|
\bigl[\frac{1}{2}(1+\cos^2\theta)(1-z^2)+\nonumber\\
&&(\alpha^2\sin^2\theta z^2+
\sqrt{2}\sin\theta\cos\theta z(1-z^2)^\frac{1}{2}\bigr]^\frac{1}{2}
\end{eqnarray}

On the other hand we obtain for heat current along a:

\begin{eqnarray} 
\frac{\kappa_{xx}}{\kappa_n}&=& 
\frac{1}{3\pi}\frac{v_a^2}{\Delta^2}(eH)I_C(\theta)F_C^{xx}(\theta,\phi)
\ln^2\bigl(\frac{\Delta}{\tilde{v}\sqrt{eH}}\bigr)\\
F_C^{xx}(\theta,\phi)&=&F_B^{xx}(\theta,\phi)
+\sqrt{2}(1+\cos^2\theta)^\frac{1}{2}\nonumber
\end{eqnarray}

As is readily seen $\kappa_{zz}$ depends only on $\theta$, while $\kappa_{xx}$
depends both on $\theta$ and $\phi$. Both angular dependences are
shown in Fig.~\ref{CTHETA} and Fig.~\ref{CPHI} respectively.
 
In both B and C phase $\kappa_{ij}$ (ij= xx,zz) and also the specific
heat which is determined by I$_{B,C}$ exhibit clear cusps at the poles
$\theta$=0 (and $\theta=\pi$) caused by the contributions from the
respective second
order node points which are present in both B and C phase. This is very
similar to what has been recently observed and analyzed in
YNi$_2$B$_2$C\cite{Izawa02b}. There the second order node points lie
along the equator and hence the
cusps appear as function of $\phi$. The most
significant difference in the B- and C- phase results can be seen
in the behaviour of $\kappa_{zz}$ for small polar
angle $\theta$. While in the B- phase it approaches zero it remains
finite in the C- phase. Furthermore the $\kappa_{xx}-\kappa_{zz}$ anisotropy
for $\theta=\frac{\pi}{2}$ is considerably smaller in the C- phase as
compared to the B-phase. In both B- and C- phase the xx- component
exhibts nonmonotonic behaviour as function of $\theta$. 

The thermal Hall coefficient in the C- phase reads

\begin{eqnarray}
\frac{\kappa_{xy}}{\kappa_n}(\theta)&=&
-\frac{v_a^2}{3\Delta^2}(eH)I_C(\theta)F_C^{xy}(\theta,\phi)
\ln^2\bigl(\frac{\Delta}{\tilde{v}\sqrt{eH}}\bigr)
\end{eqnarray}

where F$_C^{xy}(\theta,\phi)=F_B^{xy}(\theta,\phi)$ holds
because the contributions from
perpendicular node lines with \vH lying at half angle in between
cancel and so as in the B- phase one is left with polar and
equatorial contributions to the thermal Hall constant.
For numerical calculations we used the anisotropy ratio
$\alpha=\frac{v_c}{v_a}$ =1.643. It can be directly obtained from 
experimental anisotropies of thermal and electrical
conductivities\cite{Joynt02} $\frac{\sigma_c}{\sigma_a}$=
$\frac{\sigma_c}{\sigma_a}$=2.7 which are equal to $\alpha^2$.

\section{Concluding Remarks}

We have found that in \U at low
temperatures (i.e. T$\ll\tilde{v}\sqrt{eH}$) and in the superclean limit 
($(\Gamma\Delta)^\frac{1}{2}\ll\tilde{v}\sqrt{eH}$), the thermal
conductivity exhibits clear angular dependence which will help to
identify the nodal directions in \De and to verify the predictions of
the commonly discussed E$_{2u}$ model for the gap function in B and C
phases. Most significantly we predict that i) cusps
appear in the thermal conductivity and specific heat for $\theta$=0
(and $\theta=\pi$) due to the polar point nodes of \U in both
B- and C- phase. ii) in the C- phase for heat current along the nodal
direction ($\phi=\frac{\pi}{4}$) a finite
$\kappa_{zz}(\theta,\frac{\pi}{4})$ occurs even for
$\theta\rightarrow$0 which is caused by the contribution from the
additional perpendicular node regions. For the B- phase this
contribution vanishes.
iii) the $\kappa_{xx}-\kappa_{zz}$ anistropy for \vH in the ab- plane
($\theta =\frac{\pi}{2}$) is considerably larger in the B- phase as
compared to C- phase. This is again caused by the perpendicular node
lines which contribute and enhance  $\kappa_{xx}$ only for the C-
phase.

We hope that these different features in the field- angle dependent
thermal conductivity tensor above and below the critical field of the
B-C transition will resolve a part of the remaining controversy
surrounding \De in \U. 
We recall that thermal conductivity experiments have been very useful
to identify \De in unconventional superconductors. For example Izawa
et al have succeded to identify \De in Sr$_2$RuO$_4$\cite{Izawa01a},
CeCoIn$_5$\cite{Izawa01b},
organic salts\cite{Izawa02a} and more recently
YNi$_2$B$_2$C\cite{Izawa02b}. Indeed the thermal conductivity appears
to provide a unique window to access the nodal structure in unconventional
superconductors.

\vspace{1cm}
\noindent
{\em Acknowledgement}\\ 
We would like to thank Koichi Izawa and Yuji Matsuda for useful
discussions.


\end{document}